%
%

\ifx\mnmacrosloaded\undefined \input mn \fi

\input psfig


\def\Mpc{\,\hbox{Mpc}}
\def\ergs{\,\hbox{erg}\,\hbox{s}^{-1}}

\def\sqdeg{\,\hbox{deg}^2}
\def\kev{\,\hbox{keV}}

\def\ksec{\,\hbox{ksec}}
\def\kms{\,\hbox{km}\,\hbox{s}^{-1}}

\begintopmatter

  \title{A Dynamical Study of Optically Selected Distant Clusters} 

  \author{Richard~G.~Bower,$^1$ F.~J.~Castander,$^2$ Warrick~J.~Couch,$^3$ 
	R.~S.~Ellis$^2$ and H.~B\"ohringer$^4$} 

   \affiliation{$^1$ Dept.\ of Physics, University of Durham, South Road, Durham, 
	DH1 3LE, U.~K.}
   \affiliation{$^2$ Institute of Astronomy, Cambridge University, Madingley Road,
	Cambridge, CB3 0HA, U.~K.}
   \affiliation{$^3$ School of Physics, University of New South Wales,
	NSW 2052, Australia}
   \affiliation{$^4$ Max Planck Institut f\"ur Extraterrestrische Physik,
	85740 Garching, M\"unchen, Germany}

  \shortauthor{R.~G.~Bower et al.}
  \shorttitle{Optically Selected Clusters at z=0.4}

  \abstract{\baselineskip=12pt
We present a programme of spectroscopic observations of galaxies in a sample
of optically-selected clusters taken from the catalogue of Couch
et al (1991). Previous ROSAT observations of these clusters have shown them to
have lower X-ray luminosities, given their optical richness, than might be
expected on the basis of local samples. In the present paper we extend this
work by determining velocity dispersions of a subsample of the clusters.
We confirm the dynamical reality of all but one of the
original sample, and find velocity dispersions comparable with present-day
clusters of equivalent comoving space density. Thus, in the context of the
$L_X-\sigma$ relation for present-day clusters, there is evidence for a 
higher velocity dispersion at fixed X-ray luminosity. 

A key question is whether the high velocity
dispersions are indicative of the gravitational potential. If they are, the X-ray 
luminosities measured in Bower et al., 1994 (Paper~I), would then imply an 
implausibly low efficiency of X-ray generation. Alternatively, the discrepancy 
could be explained if the clusters were systems of lower virial temperature, 
in which the apparent velocity dispersion is inflated by an infalling, 
unrelaxed halo. By co-adding our
sample, we are able to consider multi-component fits to the velocity distribution
and to demonstrate evidence for a large infalling population. This might result
either from an increase with redshift in the infall rate for
clusters, or from the preferential selection of clusters embedded in 
filaments oriented along the line of sight. Since clusters with similar
properties can be found in local optically selected catalogues, we suggest
that the latter explanation is more likely.
	}

\maketitle

\section{Introduction}

This work is based on the catalogue of optically-selected distant clusters compiled by
Couch et al., 1991 (CEMM). Clusters were selected purely on the basis
of the over-density of faint galaxies on contrast enhanced AAT 4-m 
photographic plates. The selection was not implemented by computer 
algorithm, but objectivity was  tested exhaustively using simulations. 
The total area of sky searched was $46\sqdeg$.
Clusters with a density enhancement more than $4\sigma$ above the local
background were then selected for spectroscopic observation  
in order to determine their redshifts. The redshift of a cluster was 
only accepted when two or more galaxies with consistent galaxies were found. 

Bower et al. (1994, here after Paper~I) first used this catalogue to study the
X-ray evolution of clusters. A subset of 14 clusters were targeted in pointed
ROSAT observations. The clusters were found to have surprisingly low X-ray 
luminosities, indeed several were not detected at all despite exposure times 
in excess of $14\ksec$. It was shown that this result could be explained 
by mild negative evolution of the X-ray luminosity function of the type 
suggested by Henry et al.\ (1992) on the basis of the initial EMSS survey.
Similar results have been found in other surveys based on optical 
selection criterion (eg., Nichol et al., 1994, Holden et al., 1997),
although the most recent X-ray selected surveys conflict over the extent of
the evolution they find (eg., Castander et al., 1995, Rosati et al., 1995,
Nichol et al., 1997, Collins et al., 1997).

In this paper, we present a dynamical study of the CEMM cluster sample.
Initially, the data will be used to confirm the physical 
existence of the clusters that we have targeted. The cluster members
will then be used to investigate the velocity structure of the clusters.
In particular, we will compare the velocity dispersion with the 
cluster X-ray luminosities presented in Paper~I, using the present-day
X-ray luminosity -- velocity dispersion ($L_X$--$\sigma$) correlation as 
a guide. There are three physical regimes we can consider. 
Firstly, if the evolution of the clusters is driven primarily by the 
evolution of their gravitational structure, we should expect their 
velocity dispersions to be smaller than those of present-day 
clusters of equivalent luminosity (eg., Kaiser, 1991). Secondly, evolutionary models in which
the entropy of the cluster gas is a dominant factor predict that the
evolution of cluster luminosities and velocity dispersions will be parallel
each other so that little evolution of the $L_X$--$\sigma$ is  
expected (Henry \& Evrard, 1991). The third possibility is that the velocity dispersions are
higher than can be explained by the models discussed above.  
Such a situation could result if the dynamical state of the clusters
was dominated by infall and/or merging. In this case, the velocity
dispersion would no longer be representative of the virial temperature
of the cluster (eg., Frenk et al., 1990). 

As we will show in the following sections, it is difficult to obtain
redshifts for large samples of galaxies in these clusters, and we must
consider how many redshifts are necessary to determine an adequate
estimate of the velocity dispersion. In an isolated cluster, a 
sample of 10 cluster members, are relatively accurate measurement can, 
theoretically, be obtained. For example, a measured dispersion of
$600 \kms$ might arise from a system with true velocity 
dispersion of between $498\kms$ and $812\kms$ at the 1$\sigma$ (68\%) 
confidence level. This accuracy is adequate to distinguish between massive 
and poor clusters (eg., Zabludoff et al., 1993). 
To obtain better definition, we would have to substantially increase the 
number of measured redshifts. In practice, however, although adding further 
members increases the measurement precision only slowly, it greatly reduces 
the sensitivity of the estimated velocity dispersion to inclusion or 
exclusion of outlying galaxies. Unfortunately it is extremely inefficient to 
increase the size of datasets for the clusters individually, so
we address this problem by combining the individual clusters to create
a single synthetic system: sufficient data are then available to robustly 
estimate the average dispersion. Furthermore, two component fitting can be 
used to investigate whether the clusters can be separated
into a virialised core and higher dispersion infalling halo.

The outline of the paper is as follows. Section~2 sets out the data
on which this paper is based. Specifically, Section~2.1 summarises the X-ray data
from Paper~I and presents data for the additional cluster F1557.19TC
($z=0.51$), while Section~2.2 describes the spectroscopic data that are
central to this study. The analysis of the cluster redshift distributions
is detailed in Section~3, including our analysis of the combined dataset. 
In Section 4, we discuss the implications of the high dispersions that
we find, including a comparison with evolutionary models. Our conclusions
are presented in Section~5.

\section{Data Reduction}

\subsection{X-ray Data}

The clusters studied in this paper were selected from the CEMM
clusters for which we had obtained pointed ROSAT PSPC observations, the 
seven clusters being chosen in such a way as to optimise the efficiency of the
spectroscopic observing programme.

\begintable{1}
{

      \newdimen\digitwidth
      \setbox0=\hbox{\rm0}
      \digitwidth=\wd0
      \catcode`~=\active
      \def~{\kern\digitwidth}

      \def\d{\rlap{$^\dagger$}}      
    
      \def\mkrule{\noalign{\vskip 5pt}\noalign{\hrule}\noalign{\vskip 10pt}}

      \def\note#1{\item{$^{#1}$}}

    \noindent{{\bf Table 1.}
	X-Ray Properties of the Cluster Sample}

   \bigskip

 \tabskip 1em plus 2em minus 1em 
 \halign to \hsize{
   #\hfil&    \hfil#&  \hfil#&  \hfil#& \hfil$#$& \hfil#& \hfil#\quad\cr
   Cluster&  Redshift& N$_{\rm H}$$^1$&  Exposure$^2$& 
				\hbox{Cts$^3$}& Backgr$^4$& 
						\omit\hfil Luminosity$^5$\cr
   \mkrule
   F1557.19TC&  0.51~&  4.0&   21.03~&  36.~&   94.~&  0.477\cr  
   F1652.20CR&  0.41~&  3.2&   15.10~&  34.~&   66.~&  0.395\cr
   F1637.23TL&  0.48~&  1.1&   23.98~&  48.~\d& 39.~&  0.487\cr 
   J2175.15TR&  0.41~&  1.4&   11.50~&  45.~&   26.~&  0.647\cr
   J2175.23C&   0.40~&  1.4&   19.77~&  20.~&   53.~&  0.156\cr
   F1835.22CR&  0.469&  3.9&   21.30~&  19.~\d& 42.~&  0.225\cr 
   F1835.2CL&   0.377&  3.9&   17.20~&  -8.~\d& 37.~&  $< 0.142$\cr
   \mkrule 
  }

\bigskip


\font\small = cmr8
\small

\noindent  Notes:
  \note{1} Hydrogen column density towards cluster in units of 10$^{20}$ atoms cm$^{-2}$.
  \note{2} Exposure time in ksec, corrected for telescope vignetting. 
  \note{3} Photon count in detection cell after background subtraction. Where marked $\d$\ ,
                the photon count is measured in channels 52--201, otherwise in channels 41--240. 
                For pointed data, a detect cell size of 3$'\times$3$'$ has been used; All Sky Survey data uses
                a larger detect cell (4$'\times$4$'$) to allow for the wider point-spread function.
  \note{4} Background count rate determined from spline fit (pointed observations) or mean of surrounding area (All Sky 
                Survey observations).
  \note{5} Cluster luminosity in the 0.7--3.5 KeV (in the cluster rest-frame)
		in units of 10$^{44}$ erg s$^{-1}$. A factor of 0.7 is included in order to allow for
		the flux that is lost from the detected cell.
                Where the cluster is not detected with greater than 99\% confidance, the flux 
		corresponding to the 99\% detection threshold is given.

 }
\endtable

The X-ray data for these systems have for the most part been presented
in Paper~I. The exception is the cluster F1557.19TC, for which the
data were not presented because its redshift of 0.51 exceeds the
completeness limit that we set in the earlier paper.
The X-ray data for F1557.19TC were reduced in identical manner
to that used in Paper~I. To summarise, we measure the total 
(ie., cluster plus background) flux falling within a $3'\times3'$ detect
cell centred on the optical position of the cluster. The flux was then
compared with the background flux measured from a background spline
fit. Where the excess flux in the cluster detect cell did not exceed 
the 99\% confidence detection threshold (determined by the shot noise
in the background within the detect cell), the detection threshold
is quoted as an upper limit. The X-ray flux derived in this way has been 
converted to rest-frame luminosity assuming a cluster temperature of 
$5\kev$ (throughout the paper, we adopt the
cosmological parameters $H_0 = 50 \kms\Mpc^{-1}$ and $q_0 = 0.5$).
The cluster luminosities are presented in Table~1.

Due to an error in Paper~I, the X-ray luminosities given there were 
incorrectly quoted as referring to the cluster rest-frame 0.5--$2.5\kev$ 
energy band. In actuality, the table gave luminosities in the observed-frame
energy band. For consistency, the data in Table~1 have 
been referenced to the 0.7--$3.5\kev$ energy band in the cluster rest-frame. 
For a cluster with redshift 0.40, the two energy bands are equivalent. 
Small differences from the values presented in Table~1 of Paper~I
occur when the actual redshift of the cluster is taken into account.
Correcting the error in Paper~I shifts the best fitting luminosity function 
by $-0.15$ in $\log(\hbox{luminosity})$, further increasing the significance 
of the result presented there.

\subsection{Spectroscopic Data}

Spectroscopic observations were undertaken to considerably supplement the
redshift data obtained for the clusters by CEMM. The observations were 
made with the 3.6m telescope at the European Southern Observatory, La Silla,
using the high throughput EFOSC grism spectrograph (D'Odorico, 1990, Melnick, 1991). 
The data were gathered during two observing runs in January and December, 1994:
the clusters F1835.22CR, F1835.2CL and J2175.23C being targeted during the
first run; F1557.19TC, F1652.20CR, F1637.23TL and J2175.15TR in the second. 
Both runs used the same instrumental set-up, the O150 Grism (and Tektronix CCD \#26)
providing a dispersion of 3.6\AA\ per pixel and an empirical resolution of 
10\AA\ (FWHM). 
 
In order to achieve our target of at least 10 members in each cluster, we observed
each cluster twice using multi-object masks created with the PUMA
punching machine. This enabled 24 objects to be observed in, at most,
a $3.6'$ diameter field. In practice, almost all of the cluster
members were drawn from within $0.5\Mpc$ of the cluster centre given by
CEMM.  With the aim of reducing field contamination, we attempted to 
preferentially select red objects for observation
(using colours estimated from the original photographic plates and
short exposure CCD frames), but found that the constraints of requiring both
wavelength coverage from at least 3600\AA\ to 4500\AA\ in the cluster
rest frame (which determines the acceptable range of spatial offsets
from the masks center), and a minimum slit length of $10''$ allowed 
relatively little freedom in the choice of objects. 


\begintable*{2}
{


      \newdimen\digitwidth
      \setbox0=\hbox{\rm0}
      \digitwidth=\wd0
      \catcode`~=\active
      \def~{\kern\digitwidth}

	\def\-{\hbox{\ ---\ }}
	\def\:{\rlap{:}}
        \def\+{$\dag$}

	\parindent=0.75cm
        \parskip=0.1cm
	\rightskip=-1.5cm
        \par


      \tabskip 0.2em plus 2em minus 1em 

      \def\mkrule{\noalign{\vskip 5pt}\noalign{\hrule}\noalign{\vskip 10pt}}

\noindent{\bf Table 2.} New Spectroscopic Redshifts

\medskip
\hbox{
  \vbox to 8cm{ 
    \halign{#\hfil&#\hfil&#\hfil&
	\hskip5mm#\hfil&#\hfil&#\hfil\cr
\multispan6{\hfill{\bf Table 2a.}  F1557.19TC\hfill}\cr
  Obj&  ~~z&	&	   Obj&   ~~z&	\cr		
  \mkrule
  A2~~~&  0.245~~& e&           B2~~~&  0.539~~\cr
  A4&  M star&   &	      B5&  0.507\cr
  A5&  0.539&    e&           B6&  0.508\cr
  A6&  0.509&    &            B9&  0.510&    \cr
  A7&  0.185&    e&           B13&  0.426&    e\cr
  A8&  0.507&    &            \cr
  A9&  0.517&    &            \cr
  A10&  0.509&    &           \cr
  A11&  0.512&  e\+&          \cr
  A12&  M star&   &           \cr
  \mkrule
    }
  \vfil}	
  \hskip5mm
  \vbox to 8cm{
    \halign{#\hfil&#\hfil&#\hfil&
	\hskip5mm#\hfil&#\hfil&#\hfil\cr
\multispan6{\hfill{\bf Table 2b.}  F1652.20CR\hfill}\cr
  Obj&  ~~z&	&	   Obj&   ~~z&	\cr		
  \mkrule
   A1~~~&  0.353~~&   e&       B1~~~&0.353~~&   e\cr
   A2&  0.409&   &           B3&  0.411\cr
   A3&  0.411&   &           B4&  0.077&   e\cr 
   A4&  0.410&   &           B5&  0.212& e\+\cr
   A5&  0.406& e\+&          B6&  0.420\cr
   A6&  0.411& e\+&          B9&  0.217& \cr
   A7&  0.255&  e&          \cr  
   A8&  0.417&   &          \cr
   A9&  0.409&   &          \cr
  A11&  0.478&   &\cr
  A12&  0.218&   e&\cr
  A14&  0.482&   e&\cr
  A15&  0.411&   e&\cr
  A16&  M star\cr
  \mkrule
    }
  \vfill}	
\hskip 5mm
  \vbox to 8cm{ 
    \halign{#\hfil&#\hfil&#\hfil&
	\hskip5mm#\hfil&#\hfil&#\hfil\cr
\multispan6{\hfill{\bf Table 2c.}  F1637.23TL\hfill}\cr
  Obj&  ~~z&	&	   Obj&   ~~z&	\cr		
  \mkrule
  A1~~~&  0.466~~& &        B5~~~& 0.478~~&   \cr  
  A2&  0.654& e\+&        B6&  0.482& e\+\cr
  A3&  0.465&    &        B10&  0.402& e\cr
  A4&  0.537& e\+&        B12&  0.478&   \cr
  A5&  0.655&   e&        B14&  0.475\cr
  A8&  0.482&   &         B15&  0.344&  e\+\cr
  A9&  0.536&    &  \cr        
 A14&  0.479&    &  \cr     
 A15&  0.810& e\+&  \cr   
 A16&  0.475& e\+&  \cr
 A19&  M star\cr
  \mkrule
    }
  \vfil}	
}	
%
%
\vskip-1.5cm
\hbox{
  \vbox to 8cm{
    \halign{#\hfil&#\hfil&#\hfil&
	\hskip5mm#\hfil&#\hfil&#\hfil\cr
\multispan6{\hfill{\bf Table 2d.}  J2175.15TR\hfill}\cr
  Obj&  ~~z&	&	   Obj&   ~~z&	\cr		
  \mkrule
 A2~~~&  0.395~~&&       B1~~~&  0.332~~&    e\cr
 A3&  0.553&   &       B5&  M star\cr
 A4&  0.553&   &       B7&  0.394\cr
 A5&  0.418& e\+ 
		  &    B8&  0.395\cr
 A6&  0.167&   e&      B10&  0.400\cr
 A7&  0.397&   &       B12&  0.331&    e\cr
 A9&  0.362&   e&      \cr
 A10&  0.395&   &      \cr
 A11&  0.394&   &      \cr
 A12&  0.392&   &      \cr
 A13&  0.337&   e\cr
  \mkrule
    }
   \vfil }	
  \hskip5mm
  \vbox to 8cm{ 
    \halign{#\hfil&#\hfil&#\hfil&
	\hskip5mm#\hfil&#\hfil&#\hfil\cr
\multispan6{\hfill{\bf Table 2e.}  J2175.23C\hfill}\cr
  Obj&  ~~z&	&	   Obj&   ~~z&	\cr		
  \mkrule
  A2b& 0.354~~&   e&        B1~~~&  0.393~~\cr
  A3~~~&  0.402&   &        B2&  0.151&	e\cr
  A4&  0.402& e\+&          B3&  0.402\cr
  A5&  0.453&   e& 	
			    B4&  0.386&     e\+\cr
  A6b& 0.373& e\+&          B5&  0.535&     e\+\cr
  A7&  0.396& e\+&          B9&  M star\cr
  A8b& 0.408&    &           B11&  0.596\cr  
  A11&  0.406&   &           B12a& 0.5&     e\+\cr  
    &    &      &            B12b& \-\cr
    &    &      &            B13&  0.471&   e\+\cr
  \mkrule
    }
  \vfil}	
  \hskip5mm
  \vbox to 8cm{
    \halign{#\hfil&#\hfil&#\hfil&
	\hskip5mm#\hfil&#\hfil&#\hfil\cr
\multispan6{\hfill{\bf Table 2f.}  F1835.22CR\hfill}\cr
  Obj&  ~~z&	&	   Obj&   ~~z&	\cr		
  \mkrule
  A1~~~&  M star&  &      B1~~~&  0.240~~&    e\cr
  A3&  M star&  &         B3&  0.473&	e\cr
  A4&  0.468~~&   &       B6&  0.500\cr
  A5&  0.471&   &         B10&  0.715&	e\+\cr
  A6&  M star&  &         B11&  0.321&	e\cr
  A7b& 0.470&   &         \cr
  A9&  0.472&   e&        \cr
  A10&  0.478&    &       \cr
  A11&  0.186&    &       \cr
  A12&  0.466&   &        \cr
  \mkrule
    }
    \vfil}	
} 
\vskip -2.2cm
\hbox{
  \vbox to 8cm{ 
    \halign{#\hfil&#\hfil&#\hfil&
	\hskip5mm#\hfil&#\hfil&#\hfil\cr
\multispan6{\hfill{\bf Table 2g.}  F1835.2CL\hfill}\cr
  Obj&  ~~z&	&	   Obj&   ~~z&	\cr		
  \mkrule
  A1a& 0.442~~&   &           B7~~~&  0.302~~&	e\cr
  A2~~~&  0.477&    &         B8&  0.486&       e\cr
  A3&  0.442&   &             \cr
  A4&  0.212&   e&            \cr
  A6a& 0.444&   e&            \cr
  A7&  0.303&   &             \cr
  A8&  0.315&   &             \cr
  A9b& 0.377&   &             \cr
  A10b& 0.415& &              \cr
  \mkrule
    }
  \vfil}
}	

\vskip -3.0cm
  \noindent{Notes:}
  \item{e} Spectrum contains emission lines with greater than 7\AA\ equivalent width.
  \item{\+} Redshift is based on only one line.

 }
\endtable

\beginfigure*{1}
\psfig{figure=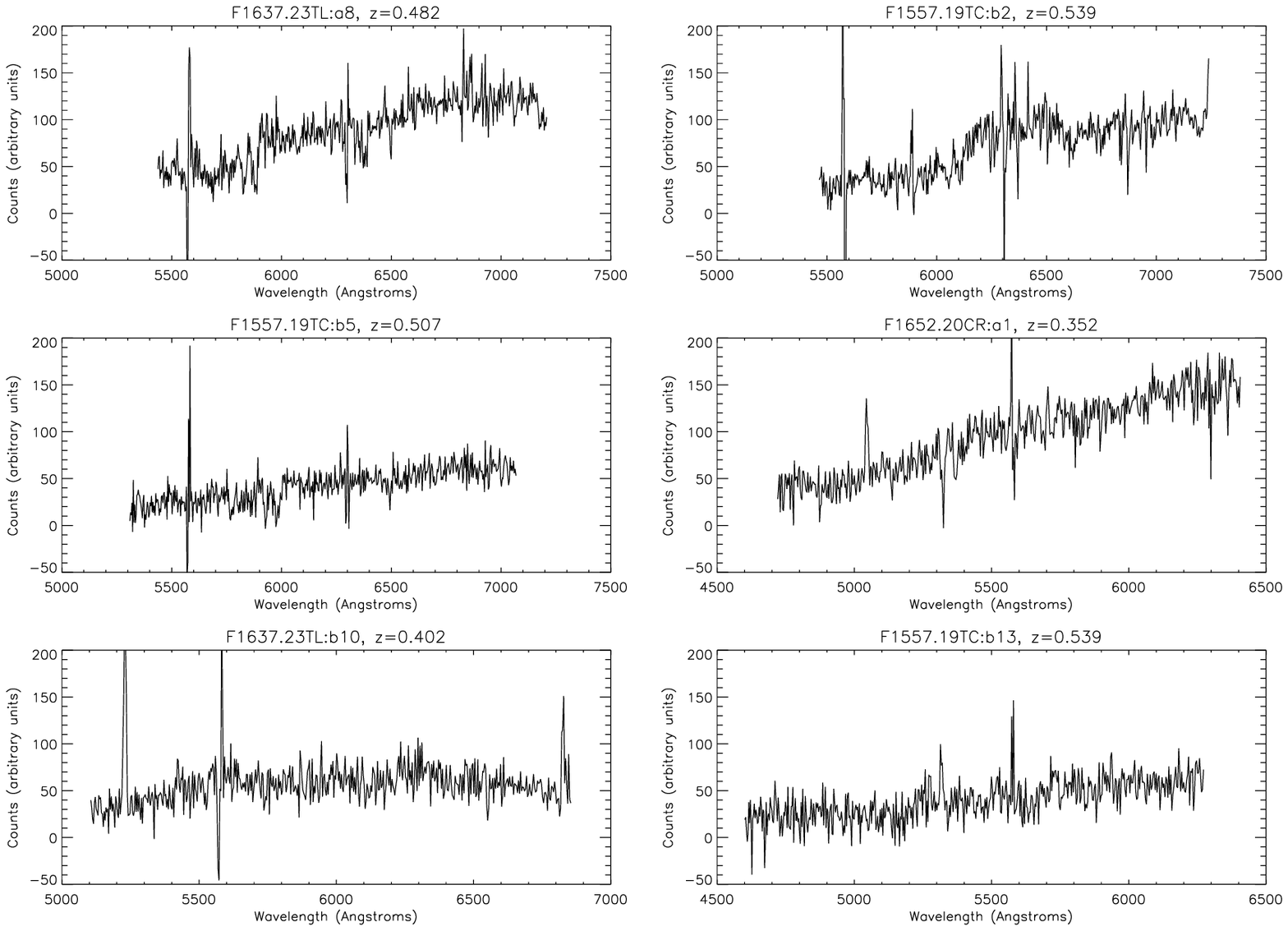,width=16.8cm}
\caption{{\bf Figure 1.} Examples of typical spectra taken with the EFOSC 
	system. These are raw spectra, and the residuals remaining for 
	subtraction of strong night-sky emission lines can be clearly seen. 
	The plots are labelled with the object code given in Table~2.}
\endfigure

The redshifts measured from our EFOSC (1994) spectra are given in Table~2,
and typical spectra are illustrated in Figure~1.
For absorption-line galaxies, redshifts were measured by overlaying a 
high signal-to-noise template spectrum constructed by co-adding the spectra
of the brighter objects. The redshift determination could then be made
by sliding the template spectrum in redshift until the strong features
matched those in the target galaxy. By offsetting the template,
the accuracy of the redshift measurement could be judged. As a result of the
high dispersion and oversampling of the spectra, we found that
once the target spectrum showed convincing absorption features, its redshift 
could be determined to better than 0.001. For a galaxy cluster at redshift 
0.4, this corresponds to a velocity resolution of $214\kms$ in the
cluster rest-frame. In order to avoid confusion by residuals from the
background subtraction, a night sky spectrum was also overlayed in this
fitting process. In some of the clusters, this is very important because
one of the strong Ca~H and~K lines can be obliterated by the night sky
emission feature at 5577\AA. 

For emission line galaxies, the redshift was measured by centroiding the 
emission feature. Where both absorption and emission features were present,
the two methods agreed to better than $\Delta z = 0.001$. In a number 
of cases, only one emission line and no discernible absorption features 
were present. Since the emission line was narrow, we assumed this was [OII] 
at 3727\AA. These redshifts are marked in Table~2 to indicate this
uncertainty. For the 
purposes of measuring the velocity dispersion of the cluster, the 
ambiguity of these measurements is unimportant since it 
principally involves field galaxies at redshifts significantly higher
or lower than the cluster. Mis-identification of the emission line is most
unlikely to result in the false assignment of a field galaxy to the 
cluster. 

Galaxies with emission-line equivalent width greater than 
7\AA\ are noted in Table~2. This division is used to investigate the
dependence of the measured velocity dispersion on star formation rate
in Section~3.

\begintable{3}
{

      \newdimen\digitwidth
      \setbox0=\hbox{\rm0}
      \digitwidth=\wd0
      \catcode`~=\active
      \def~{\kern\digitwidth}

	\def\-{\hbox{\ ---\ }}
	\def\:{\rlap{:}}
\nopagenumbers

      \def\mkrule{\noalign{\vskip 5pt}\noalign{\hrule}\noalign{\vskip 10pt}}

      \def\note#1{\item{$^{#1}$}}

\noindent{\bf Table 3.}  Revised Redshifts for CEMM Spectra

$$\vbox{ 
\halign{#\hfil&\qquad#\hfil&\qquad#\hfil&\quad#\hfil&
		\qquad#\hfil&\qquad#\hfil\cr
 Cluster&     Obj&   ~~z&  comment\cr	
 \mkrule
 F1557.19TC&     	a&	0.509\cr 
	&	c&	0.509\cr
 \noalign{\vskip5pt}         
 F1652.20CR&	\multispan3{\it\qquad CEMM spectra not available}\hfil\cr
 \noalign{\vskip5pt}
 F1637.23TL&    	1&	0.479\cr
	&	4&	0.465&	OII emission\cr
	&	5&	0.482\cr
 \noalign{\vskip5pt}
 J2175.15TR&     	a&	0.396\cr
	&	d&	0.406\:\cr
	&	j&	\-\cr
 \noalign{\vskip5pt}
 J2175.23C&     	a&	0.405\cr
	&	b&	0.407\cr
	&	c&	0.406\cr
 \noalign{\vskip5pt}
 F1835.22CR&    	\multispan3{\it\qquad CEMM spectra not available}\hfil\cr
 \mkrule
}
}$$

 }
\endtable

Finally, we returned to the longslit spectra obtained by CEMM (where available)
to provide redshifts for an additional 2--3 cluster members in each system. 
As CEMM required the redshifts only to establish cluster membership,
we considered it necessary to re-reduce the spectra and re-measure the galaxy
redshifts. Revised redshifts were determined using
the techniques and templates discussed above, with the sky spectra from
the two data sets being monitored to check the consistency of the data 
reduction. The reworked redshifts are given in Table~3. In the few cases 
where the measurements differ, our more precise measurements are to be preferred. 

\section{Results}

\subsection{Redshift Distribution Histograms}

\beginfigure*{2}
\psfig{figure=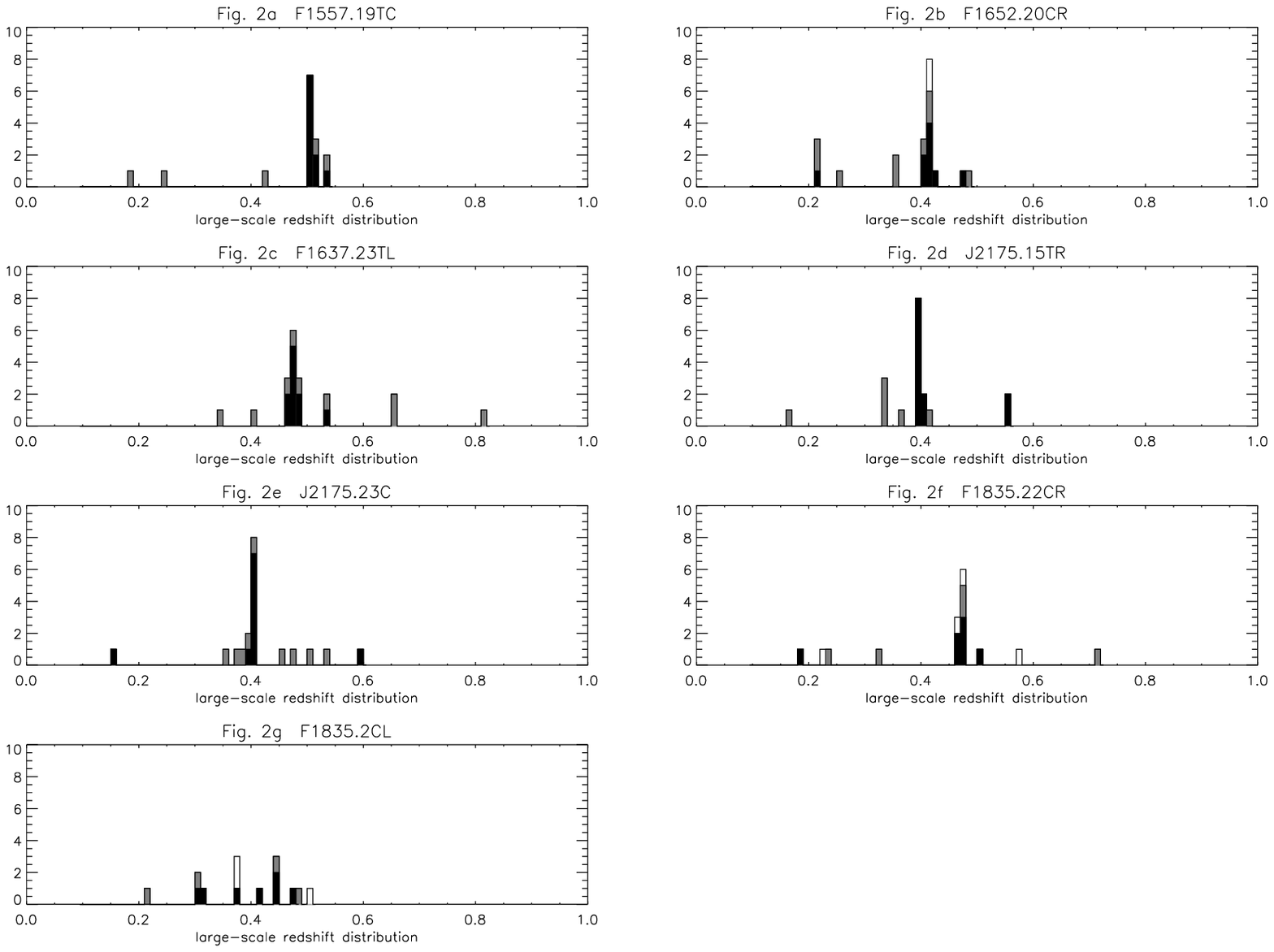,width=16.8cm}
\caption{{\bf Figure 2(a--g).} The large-scale redshift distribution of all 
	objects for which velocities have been measured. The
	shading identifies the different galaxy types: solid, absorption line
	galaxies; half-tone, emission line objects. Redshifts taken from CEMM's
	paper but which were unavailable for re-reduction are left unshaded.
	In all but the case of F1835.2CL, there is a single clear density peak
 	in redshift space.}
\endfigure

\beginfigure*{3}
\psfig{figure=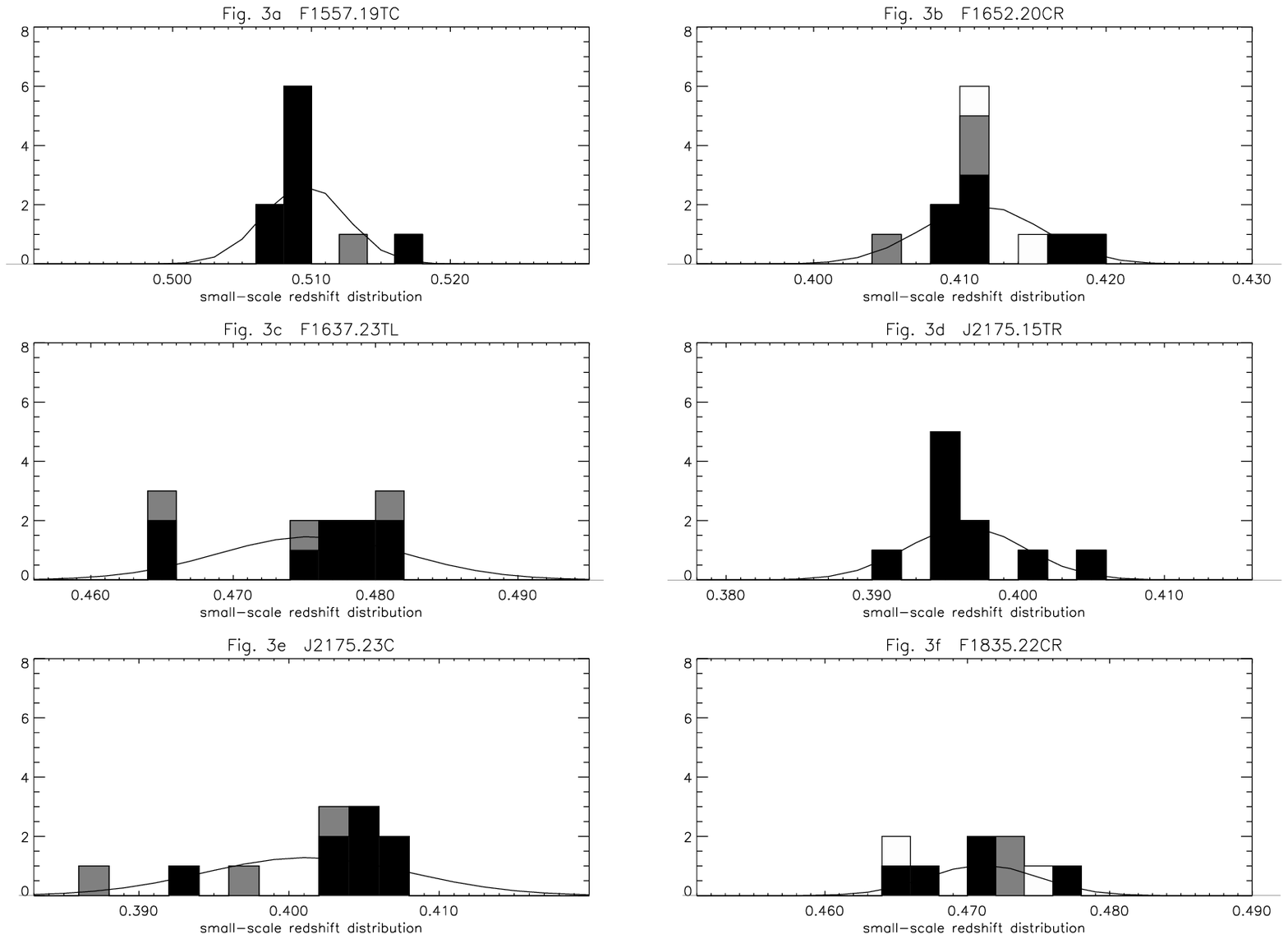,width=16.8cm}
\caption{{\bf Figure 3(a--f).} The small scale redshift distribution centered 
	on each of the bona-fide CEMM clusters. The shading of the histogram 
	distinguishes the object types as in Figure~2. The Gaussian 
	distribution overlaid shows
	the Method~1 fit given in Table~4. All plots have been scaled to show a
	range of $\pm 4000\kms$ (in the cluster rest-frame) about the mean
        cluster redshift.}
\endfigure 

Figures~2(a--g) and 3(a--f) present the velocity histograms for each cluster. 
Two views are given, one showing the distribution on a coarse scale 
(Figure~2a--g), the second focusing on a rest-frame velocity range of 
$\pm4000\kms$ about the cluster mean (Figure~3a--f). F1835.2CL is not shown
in Figure~3 because of the lack of any single galaxy concentration.
The first point we conclude from these figures is that, in all but one case, 
the cluster (although identified from its two dimensional projection
on the sky) corresponds to a genuine physically associated system: the
selection of the cluster by CEMM as a two dimensional density enhancement
is confirmed as a sharp spike in redshift space in our data.
The exception to this is the cluster F1835.2CL, for which the 12 galaxies
that could be measured returned only one redshift close to the cluster
redshift proposed by CEMM.

The small-scale panels allow us to visually assess the cluster velocity
dispersions. It can immediately be appreciated that while a velocity
dispersion can be readily calculated from these galaxies, it is very much
more difficult to be certain that this value is representative of the
gravitational potential of the cluster. There are essentially two kinds of 
contamination that are liable to affect our results. Firstly, the 
cluster may actually be composed of two (or more) distinct gravitational 
components. Hopefully, such a situation would be apparent from the 
double-peaked nature of the small-scale redshift distribution. If the
velocity separation of the cluster components is small, however, the
small sample of redshifts available to us is unlikely to show a statistically
significant effect. It seems unlikely, however, that this problem would
equally affect all of the clusters. Secondly, individual outliers such as
close-by field galaxies may appear close to the cluster in redshift space 
and thus artificially inflate the measured dispersion. This possibility 
must be countered by applying a variety of clipping algorithms to the
data and testing the robustness of the dispersions that we find.

\subsection{Measuring the Cluster Velocity Dispersion}

\begintable*{4}
{

      \newdimen\digitwidth
      \setbox0=\hbox{\rm0}
      \digitwidth=\wd0
      \catcode`~=\active
      \def~{\kern\digitwidth}

	\def\-{\hbox{\ ---\ }}
	\def\:{\rlap{:}}
	\def\d{\rlap{$^\dagger$}}

	\def\kms{\,\hbox{km}\,\hbox{s}^{-1}}

      \def\mkrule{\noalign{\vskip 5pt}\noalign{\hrule}\noalign{\vskip 10pt}}

      \def\note#1{\item{$^{#1}$}}

      \tabskip 1em plus 2em minus 1em

\parindent=0.75cm
\leftskip=-3.5cm
\rightskip=-4.5cm

\noindent{\bf Table 4.}  Cluster Velocity Dispersions

\baselineskip=10pt

\halign{#\hfil&\quad#\hfil&\quad\hfil#\qquad&\hfil#&
	\hfil#&\hfil#&\quad#\hfil&\quad#\hfil\cr
 Cluster&     ~~z& \omit\hfil $\sigma_v\,^a$\hfil&   
				N&  $\sigma_v^+\,^b$&   $\sigma_v^-\,^b$&
				 $P_{KS}\,^c$&  \omit\hfil Method$^d$\hfil\cr
 \mkrule
 F1557.19TC&   	0.5095&	584&	10&	791&	481&	0.23&	1\cr
	&	0.5089&	355&	9&	492&	291&	0.09&	2\cr
	&	0.509&	572&	10&	775&	474&	0.06&	3\cr
	&	0.5094&	597&	9&	827&	490&	0.07&	4\cr       
	\noalign{\smallskip} 
 F1652.20CR&    0.4115&	863&	10&	1167&	714&	0.30&	1\d\cr
	&	0.4115&	1004&	10&	1358&	831&	0.22&	2\d\cr
	&	0.411&	826&	10&	1117&	683&	0.40&  	3\d\cr
	&	0.4124&	916&	7&	1357&	736&	0.66&	4\d\cr
	\noalign{\smallskip}
 F1637.23TL&    0.4755&	1335&	12&	1744&	1121&	0.80&	1\cr
	&	0.4755&	1553&	12&	2029&	1304&	0.84&	2\cr
	&	0.478&	1373&	12&	1793&	1153&	0.17&	3\cr
	&	0.4766&	1284&	9&	1780&	1055&	0.86&	4\cr
	\noalign{\smallskip}	
 J2175.15TR&    0.3964&	854&	10&	1156&	707&	0.52&	1\cr
	&	0.3953&	559&	9&	774&	459&	0.33&	2\cr
	&	0.395&	865&	10&	1171&	716&	0.06&	3\cr
	&	0.3969&	834&	10&	1156&	707&	0.52&	4\cr
	\noalign{\smallskip}
 J2175.23C&     0.4012& 1463&   11&  	1942&	1221&	0.84&	1\cr
	&	0.4027&	1213&	10&	1641&	1004&	0.76&	2\cr
	&	0.402&	1405&	11&	1865&	1172&	0.33&	3\cr
	&	0.4036&	1027&	8&	1465&	835&	0.65&	4\cr
	\noalign{\smallskip}
 F1835.22CR&    0.4711& 784& 	7&	1162&	630&	0.90&	1\d\cr
	&	0.4711&	912&	7&	1351&	733&	0.71&	2\d\cr
	&	0.471&	727&	7&	1076&	584&	0.89&	3\d\cr
	&	0.4706&	930&	5&	1562&	723&	0.69&	4\d\cr
	\noalign{\smallskip}
 F1835.2CL&       \-&          \-\cr 
	\noalign{\smallskip}
	\noalign{\smallskip}     
 Composite&	&	930&	59&	1033&	852&	0.08&	1\d\cr
 \rlap{(all clusters)}
	&	&	854&	54&	953&	780&	0.50\rlap{$^*$}&  2\d\cr
	&	&	907&	48&	1018&	824&	0.12&	4\d\cr
	&	&	963&	59&	1070&	833&	0.0002&	5\d\cr
	&	&	966&	48&	1064&	860&	0.0010&	4,5\d\cr
	\noalign{\smallskip}
 Composite&	&	815&	47&	917&	740&	0.16&	1\d\cr
 \rlap{(F1637.23TL ommitted)}
	&	&	798&	44&	901&	722&	0.05&	2\d\cr
	&	&	916&	41&	1040&	826&	0.04&	4\d\cr
	&	& 	809&	47&	911&	735&	0.009&	5\d\cr
	&	&	829&	39&	945&	747&	0.15&	4,5\d\cr
 \mkrule
}

\font\small = cmr8
\small

{
 \noindent{Notes:}

 \item{$^a$} Velocity dispersion of cluster members given in $\kms$ in the cluster
	restframe.

 \item{$^b$} 68\% upper and lower confidence limits for the velocity dispersion, 
	$\sigma_v$, calculated from the F-ratio test. This gives an estimate of the 
	sampling uncertainty in the measured dispersion, but makes no
	allowance for possible departure of the underlying distribution 
	from Gaussian form.  

 \item{$^c$} Folded KS test probability that galaxies are drawn from a Gaussian 
	distribution with dispersion $\sigma_v$. Folding the distribution about
	the cluster mean increases the test's sensitivity to the kertosis of the
	data-set.

 \item{$^d$} Estimation Methods:
 \itemitem{1.}  3$\sigma$ clipping. Initial value of dispersion $1500\kms$. 
 \itemitem{2.}  2$\sigma$ clipping. Measured dispersions are corrected by a factor
	including correction 1.16 to allow for galaxies that are trimmed from
	the Gaussian tail. Initial value for the dispersion is $1500\kms$.
 \itemitem{3.}  3$\sigma$ clipping about median cluster redshift. Dispersion is
			calculated without applying a $\sqrt{n/(n-1)}$ bias 
			corection.
 \itemitem{4.}  Method 1 applied to absorption line galaxies only.
 \itemitem{5.}  As method 1, but using median redshift to centralise the galaxy
	clusters before coaddition.
 
 \item{$^\dagger$} Velocity dispersion has been calculated excluding 
	uncomfirmed CEMM data.

 \item{$^*$} Centre of fitted gaussian is $0.4509$. Although a sharp peak is 
	present in the velocity distribution, this is smoothed out in the
        folded KS test.
 
\par}
 }
\endtable

The velocity dispersions of the clusters are given in Table~4. 
Four values are quoted to illustrate different 
approaches determining the central value and the effect of excluding emission-line
galaxies. All values are quoted in $\kms$ and
have been transformed to the cluster rest-frame. The velocity dispersion is defined as
$\sum_{i=1}^{n} (v_i-\bar v)^2/(n-n_{df})$, where $n$ is the number of galaxies
in the sample, $n_{df}$ is the number of degrees of freedom that have been 
absorbed in defining the cluster mean. For a single cluster, $n_{df}=1$,
for the composite cluster (see Section 3.3) $n_{df}=n_{clus}$, where
$n_{clus}$ is the number of clusters being used to make up the composite.
This definition of the dispersion compensates for bias that arises when
we use the galaxy sample to determine the mean cluster redshift. 
We refer to this value as `the cluster dispersion' in what follows. 

We chose {\it a priori\/}
to give most weight to the estimate based on iterative $3\sigma$ clipping
starting from a high nominal dispersion of $1500\kms$ (Method 1). 
This scheme is preferred since the clipping corresponds to a natural
division of cluster and background galaxies, and the high initial dispersion
ensures that a wide range of data is initially considered before outliers
are selected and rejected. The final result is insensitive to both the assumed value
of the dispersion and the initial mean value. Monte-Carlo simulations
were performed to confirm that the clipping resulted in negligible bias
in the measured dispersion. Furthermore, the scatter in the values
derived from the simulations agreed well with those expected on the basis
of the F-ratio test (eg., Kendall \& Stewart, 1973). 
F-ratio 68\% confidence limits for the estimated cluster dispersions are 
therefore given in columns 5 and 6 of Table 4. The measured (sample) 
dispersions for the cluster data are over-plotted on the small scale view 
of Figure~3. In order to assess whether a single Gaussian distribution 
provides an acceptable fit to the data, and thus to test 
whether there is any detectable substructure within the clusters, we 
performed Kolmogorov-Smirnoff (KS) tests on the unbinned velocity values.
Because of the small numbers of objects involved, the data must be folded 
about the cluster mean (ie., on the values of $|v_i-\bar v|$). The 
resulting confidence of acceptance is given in Table~4. In none of the 
clusters is the departure significant.

As can be seen from the table, this approach returns relatively high values
for the dispersion. The model Gaussian distribution is shown as a continuous 
line in Figure~3. 
Visual inspection of the figure suggests that the distribution is
more centrally peaked than the Gaussian curve in the three clusters F1557.19TC,
F1652.20CR and J2175.15TR. If the clusters are taken individually, this kurtosis
is not flagged by the KS distribution test; however, the qualitative impression 
is that of a halo of galaxies surrounding a much lower dispersion
central core. The only cluster that suggests a binary structure (ie., being
composed of two distinct velocity peaks) is F1637.23TL, but again this is
not flagged by the KS test. 

Another approach to determining the velocity dispersion 
is to use the median to determine the 
central cluster redshift, and then to perform the clipping about this value.  
This makes it more likely that outlying points will be rejected since
they have less influence on the choice of cluster centre. The actual
effect on the cluster dispersions is, however, minimal. We were also
concerned that the inclusion of strong emission line galaxies
to determine the cluster dispersion might lead to artificially high
values. This was tested by repeating the analysis for the low emission line
equivalent width galaxies only. Once again, there was little change from
the dispersion measured using all the available galaxies.

In order to make a quantitative assessment of the impact of outlying points, we 
experimented with a variety of more stringent clipping algorithms. As can be
seen from Table~4, clipping the velocities used in the calculation of
the dispersion at $2\sigma$ has a significant effect on the clusters F1557.19TC
and J2175.15TR. In these clusters, this experiment confirms the
visual impression that the velocity distribution is more centrally concentrated
than would be expected for a Gaussian distribution (ie., the distribution
is leptokurtic). In order to investigate the
effect further, however, we must co-add the data for the individual clusters. 
This approach is discussed in Section~3.4.

\subsection{The $L_X$--$\sigma$ Correlation}

\beginfigure*{4}
\psfig{figure=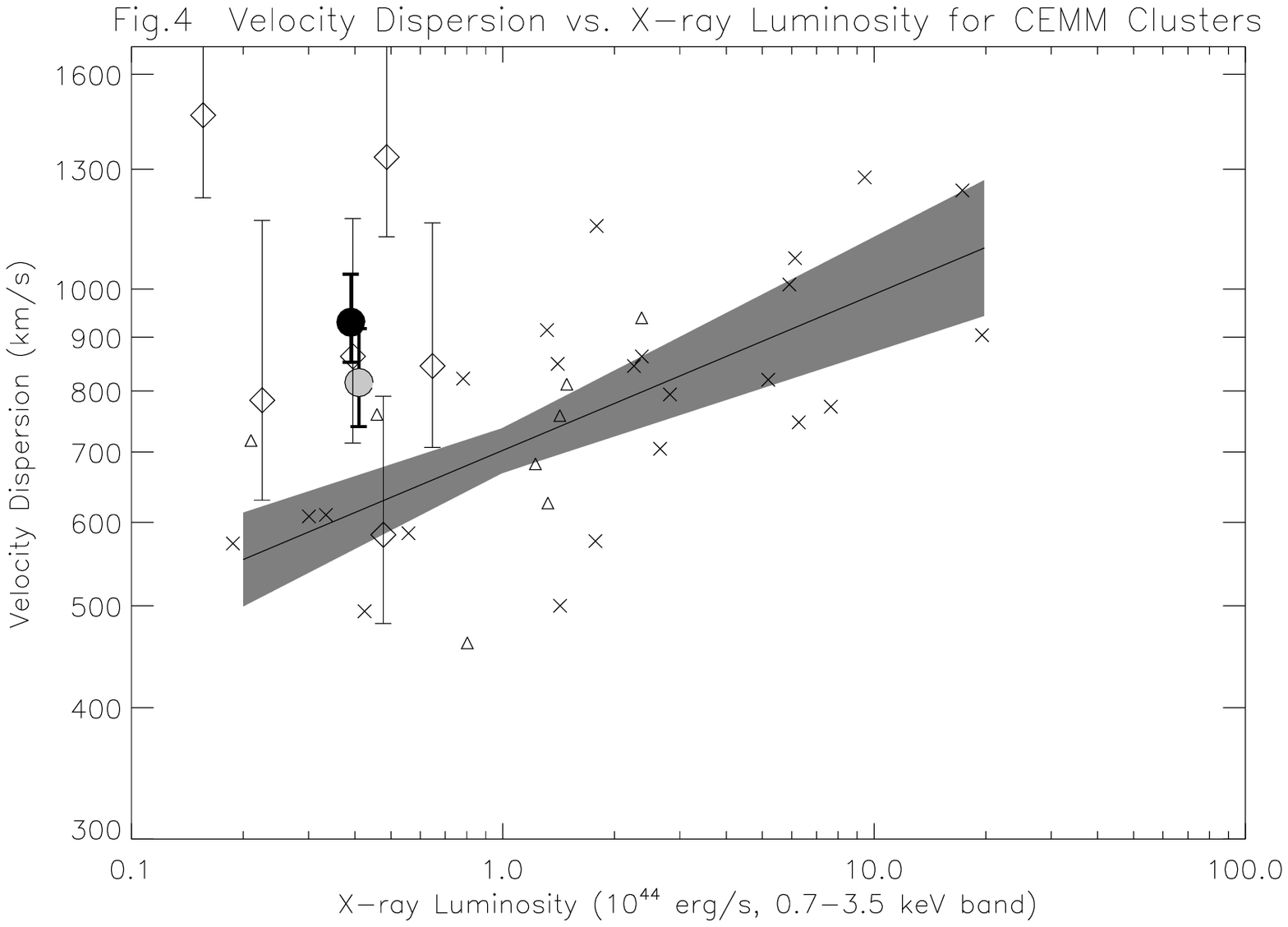,width=16.8cm}
\caption{{\bf Figure 4.} The velocity dispersions and X-ray luminosities of 
	the CEMM clusters (diamonds) compared with the present-day 
	$L_X$--$\sigma$ correlation (crosses). The dispersions of the CEMM 
	clusters plotted were determined by applying the iterative $3\sigma$ 
	clipping algorithm (Method~1 in Table~4). The position of two 
	versions of the cluster composite are shown: solid circle, all 
	clusters combined; shaded circle, omitting the cluster
	F1637.23TL. The X-ray luminosity is a simple average over the six
	individual clusters. The two points have been shifted slightly in 
	X-ray luminosity to improve clarity. The present epoch mean 
	correlation and its uncertainty are shown by the thin line and the 
	shaded region. Open triangles show the position of additional nearby 
	clusters taken from Zabuloff et al., 1990.}
\endfigure

Figure~4 compares the X-ray luminosities of the CEMM clusters with their 
line-of-sight rest-frame velocity dispersion. Data for the individual clusters
are shown as open diamond symbols. The velocity data used
are the 3$\sigma$ clipped and bias corrected values discussed above. The
error bars show the 68\% F-test confidence interval for the dispersion.
For comparison, a fiducial data-set has been taken from Edge \& Stewart 
(1991, ES)
(with luminosities and temperatures updated from David et al., 1993) 
These data are shown as crosses in Figure~4. This dataset differs from
that of CEMM in including both nearby clusters and those sampled on the 
basis of X-ray flux. We have transformed the nearby clusters 
to the 0.7--$3.5\kev$ band using the given cluster temperatures and assuming 
an iron abundance of 0.33 solar. The best fitting present-day 
$L_X$--$\sigma$ correlation is given by the thin line, with its uncertainty 
indicated by the shaded region.

Although the scatter in the relation for ES clusters is large, the velocity 
dispersions of our distant clusters lie clearly above the mean correlation 
defined by the fiducial data-set.
What cannot be determined from this diagram alone, however, is whether
the CEMM clusters are lower in their X-ray luminosity, or higher in their 
velocity dispersion. 

\subsection{Analysis of a Composite Cluster}

The dispersion measured for each individual cluster has a large statistical
uncertainty. 
For example, the cluster J2175.15TR is compatible at the $1\sigma$ level
with dispersions ranging from $707$ to $1156\kms$. Thus a
quantitative comparison with present-day clusters must be made by combining
the separate clusters. This may be achieved by averaging the results
for the individual systems, or by merging the galaxy velocities
to create a composite velocity distribution for the entire sample.
The advantage of combining our data in the second way is that outlying points,
or a non-Gaussian tail to the velocity distribution, should 
become more visible. Of course, in combining the data in this way,
two assumptions are implicit. Firstly,  
the clusters must have similar velocity dispersions. Secondly,
the systems must be assumed not to be binary (ie., composed of two distinct systems
aligned along the line of sight): if this were the case, 
co-adding the individual systems would tend to
obscure any substructure in the velocity distribution. 

\beginfigure{5}
\psfig{figure=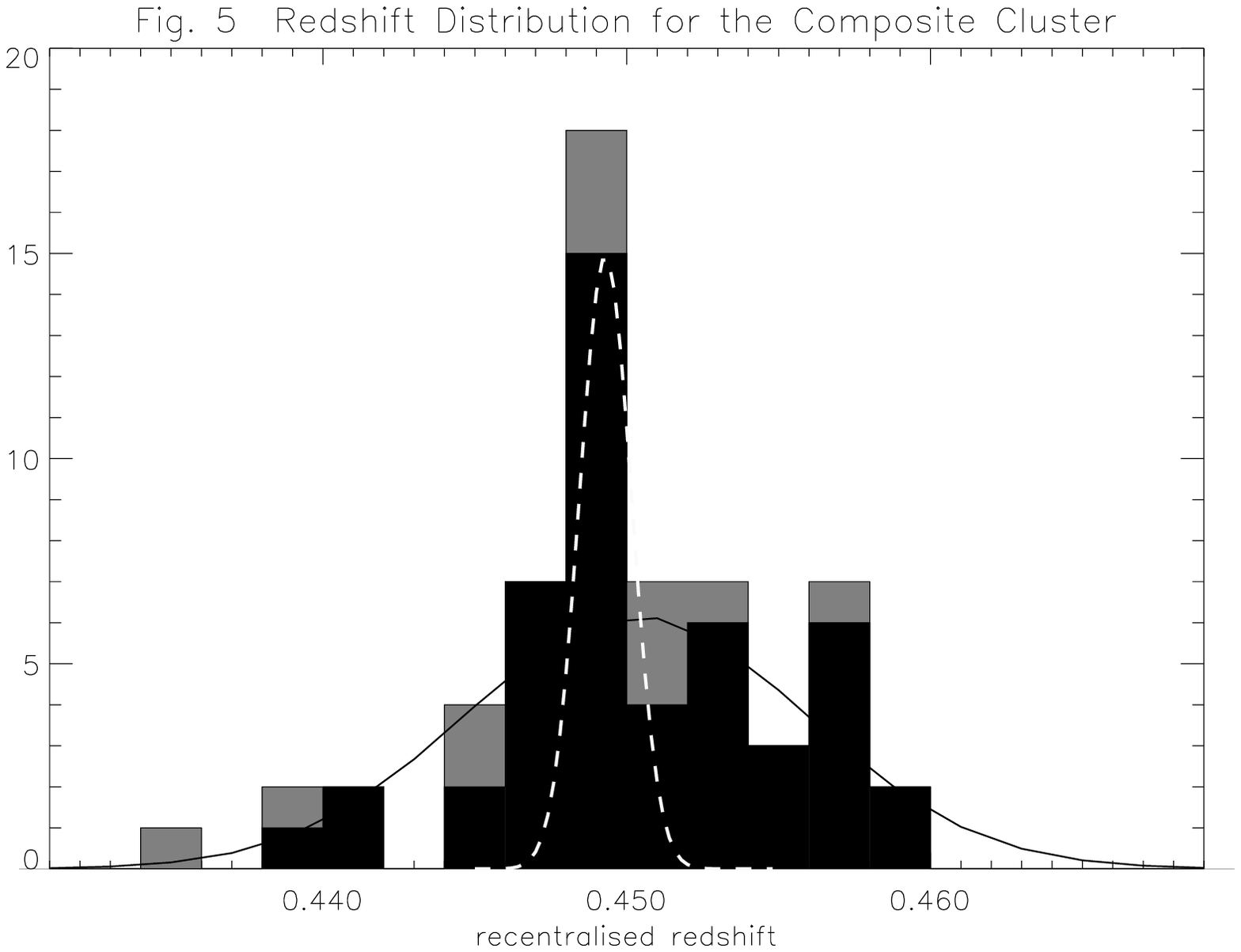,width=8.4cm}
\caption{{\bf Figure 5.} The redshift distribution of the composite cluster 
	formed by co-adding the six individual bona-fide clusters. The 
	velocity offset of each galaxy with respect to its cluster mean has 
	been converted to a redshift offset from a synthetic cluster at 
	$z=0.45$. Different shadings have been used to distinguish the 
	galaxy spectral types following Figure~2.  
	Gaussian fits derived by applying the iterative $3\sigma$ clipping
	algorithm (Method~1 in Table~4) are showing: dashed line, all galaxies;
	solid line, absorption line galaxies only.}
\endfigure

In order to combine the clusters, galaxy redshifts were converted to  
rest-frame velocity offsets about the centre of target cluster (as given in
Table~3). These were then transformed back to create a synthetic redshift
distribution for a cluster with redshift 0.45 (Figure~5). We repeated the
procedure using the median, rather than mean, to re-centralise the different
clusters prior to co-addition in order to investigate the composite's 
sensitivity to outliers. Velocity dispersions were measured using iterative
$3\sigma$ clipping, and a number of other schemes, as described in 
Section~3.2. The results are given at the end of Table~4. The measured
velocity dispersion (derived using $3\sigma$ clipping) is
$930\left(^{1033}_{~852}\right)$, significantly higher (at more 
than 99.9\% confidence) than the average value of the 
present-day $L_X$--$\sigma$ correlation. This data-point is illustrated
by a solid circle in Figure~4. 

In order to guard against the inclusion of binary clusters, the dispersion
was recalculated omitting the cluster F1637.23TL. This is the only cluster
for which the velocity dispersion histogram gives a visual impression
of binary structure. Although this structure is of low statistical 
significance, omitting this cluster tests the robustness of the final
dispersion. As can be seen from the Table~4, excluding the cluster 
slightly lowers the velocity dispersion of the composite dataset (shown as
a shaded circle in Figure~4), but the dataset is still inconsistent with 
the fiducial present-day relation at 99.5\% confidence. However, we must 
be wary of how to interp this result: the CEMM clusters have been selected 
from optical imaging, whereas the ES dataset from which we have 
constructed the present-day relation is derived from an X-ray selected
catalogue. We return to this point in Section~4. 

In addition to increasing the accuracy of the measured dispersion, 
the greater size of the composite dataset makes it possible to
test the adequacy of the single Gaussian description of the data,
and thus to check for an extended tail to the redshift distribution.
Firstly, we tested the acceptability of the fit by applying the 
Kolmogorov-Smirnoff test to the data folded about the cluster mean.
As shown in Table~4, the test rejects the single component fit with
confidence levels ranging from 85 to 99.98\% depending on how the 
composite cluster is constructed. Thus, the initial visual impression of a non-Gaussian 
distribution is confirmed in the larger dataset.

\begintable{5}
{

      \newdimen\digitwidth
      \setbox0=\hbox{\rm0}
      \digitwidth=\wd0
      \catcode`~=\active
      \def~{\kern\digitwidth}

	\def\-{\hbox{\ ---\ }}
	\def\:{\rlap{:}}
	\def\d{\rlap{$^\dagger$}}

	\def\kms{\,\hbox{km}\,\hbox{s}^{-1}}

      \def\mkrule{\noalign{\vskip 5pt}\noalign{\hrule}\noalign{\vskip 10pt}}

      \def\note#1{\item{$^{#1}$}}

\tabskip 0.5em plus 2em minus 1em

\noindent{\bf Table 5.} Maximum Likelihood Fits to Composite Distant Cluster
\medskip

\baselineskip=10pt

\halign{#\hfil&~~#\hfil&#\hfil&#\hfil&#\hfil&#\hfil\cr
 Model Fitted&\omit Galaxy\hfil&  Central&
			\omit\hfil$\sigma_v\,^b$\hfil&	Normal-&  Likelih'd\cr
 	&	\omit\hfil Sample$^a$\hfil&
			  	Redshift&($\kms$)&   ~isation&  ~Ratio$^c$\cr
 \mkrule
 single component&	A&	0.4498&	~1073&	~60.0&	\cr
	\noalign{\vskip 5pt}
 two component&		A&	0.4503&	~1169&	~43.6&	10.7\cr
		&	&	0.4493&	~~177&	~16.1\cr
	\noalign{\vskip 10pt}		
 single component&	B&	0.4497&	~1123&	~60.0&	 \cr
	\noalign{\vskip 5pt}
 two component&		B&	0.4497&	~1263&	~40.9&	26.6\cr
		&	&	0.4500&	~~~88&	~19.3\cr
	\noalign{\vskip 10pt}		
 single component&	C&	0.4498&	~~896&	~48.0&	\cr
	\noalign{\vskip 5pt}
 two component&		C&	0.4504&	~1032&	~34.1&	~8.5\cr
		&	&	0.4483&	~~175&	~13.9\cr
	\noalign{\vskip 10pt}		
 single component&	D&	0.4495&	~1052&	~48.2&	\cr
	\noalign{\vskip 5pt}
 two component&		D&	0.4507&	~1173&	~28.4&	14.8\cr
		&	&	0.4491&	~~217&	~19.6\cr
 \mkrule		
}

 \noindent{Notes:} 
  \note{a} Galaxy sample codes: A, all clusters, aligned by mean;  B, all
	 clusters, aligned by median;  C, all clusters, absorbtion line
	 galaxies only, aligned by mean;  D, as A but omitting F1637.23TL

  \note{b} Velocity dispersion is given in $\kms$ in the cluster rest frame.

  \note{c} Likelihood ratio of two component model versus single component model.

 }
\endtable

An alternative method of assessing the deviation from a Gaussian distribution
is to apply the maximum likelihood method to fit the data with a
two Gaussian components (cf., Ashman et al., 1995). 
The data were first fit with a single component in order to establish 
a reference likelihood. The minimisation adjusted both central redshift,
dispersion and normalisation. 
The resulting fit is given in Table~5: the dispersion ($1073\kms$)
is similar to that measured by the conventional approach. A second 
component was then introduced (starting from $500\kms$) and the minimisation 
procedure repeated. The new fit resolves
the cluster into core and halo components, confirming the visual impression.
The best fit dispersions of the components are $1169\kms$ and $177\kms$: the
low dispersion of the core component is striking. This fit is considerably better
than the single component version. The likelihood is improved by
$10.7$ if the clusters are aligned using their mean redshift, or
$26.6$ if aligned using the median. Interpretation of the absolute
statistical significance of these cases depends on the number of degrees of freedom
that are associated with the new component (median alignment of 
the individual clusters, in particular, will always tend to reinforce 
the impression of a core), and is best addressed by Monte-Carlo 
simulation. These show that a likelihood ratio of 11.7 is significant at the
1\% level if the clusters are aligned by their mean redshift. 
If they are aligned using the median, a higher likelihood
ratio (20.6) is needed for the same level of significance.

Thus, while the galaxies in the composite cluster have a large velocity dispersion,
there is evidence (at the 1\% level) to suggest that this may
arise from two distinct populations: a low dispersion `core' and a high 
dispersion `halo'.  Examining the separate distributions of high and low 
emission-line equivalent width galaxies suggests that the low dispersion 
component is present only in the weak-lined population. However, a KS test
does not reject the possibility that the two distributions are the same.

Care must be taken not to over-interpret the separation of the composite
distribution into `core' and `halo' components. It is clear that the low 
dispersion component accounts only for a relatively small fraction
of the total galaxy population of the cluster: less than
30\% of the total population. Therefore, a high velocity dispersion
is characteristic of the majority of the cluster galaxies, and is not the
result of a few outlying data points. Furthermore, the dispersion of the
central component is unreasonably low --- it is unresolved in our data ---
if it is indicative of the clusters' virial mass. It is almost certainly
more correct to interpret the core/halo split only as a way of characterising
the kurtosis of the velocity distribution, rather than assigning 
a definite physical significance to the two components.

\section{Discussion}

In Section~3.3, we showed that the velocity dispersions we measure for
these distant clusters lie considerably above that expected on the basis
of our fiducial $L_X$--$\sigma$ correlation. This departure can be 
interpreted in one of two ways. The first possibility is that the cluster's 
X-ray luminosity is low compared to the galaxy velocity dispersion. We
could understand this as being due to the evolution of X-ray emission from 
distant clusters (Paper~I).  The second possibility is that the discrepancy 
lies with the velocity dispersion --- perhaps a high rate of infall into 
the distant clusters has caused us to significantly over-estimate the 
virial temperature. 
Below, we consider the case for these two interpretations of the data 
in detail, and investigate the roles played by evolution and the clusters' 
optical selection.

\subsection{Comparison with Nearby Clusters and Models for Cluster Evolution}
In order to set an initial expectation of the cluster's velocity dispersion,
we use the CEMM cluster space density. This was derived in Paper~I from 
the area of sky covered by the CEMM survey, and the completeness limits 
of the catalogue. The average density of the clusters relevant to this 
work is $5 \times 10^{-7} \Mpc^{-3}$. 
(We have deliberately avoided making a comparison on the basis of the 
cluster's apparent richness since (1)~our imaging frames are too small 
to allow an accurate assessment of the background galaxy density, and 
(2)~the apparent richness of the clusters may be substantially affected 
by changes in the stellar populations of the cluster galaxies 
--- an effect that is discussed extensively in CEMM). 
Using the velocity dispersion distribution function from Zabludoff et al.\ (1993),
we can
convert between the number density of present-day clusters and their expected 
velocity dispersion. Nearby clusters of the same comoving space-density as 
the CEMM sample have a characteristic velocity dispersion of 
$750\, (\pm 50)\kms$: in terms of their space density, the clusters are 
roughly equivalent to Abell clusters at the lower end of richness class 1
(paper~1). We should make it clear that this is the expected average 
velocity dispersion: the distribution is very strongly skewed to higher 
dispersions, the tail possibly extending to dispersions as high as 
$1100\kms$ (adopting a space density a factor of 10 lower for the most 
extreme cluster).  

By contrast, the velocity dispersions that would be inferred from comparison
with present-day clusters of comparable X-ray luminosity (ie., 
$0.4\times 10^{44} \ergs$ in the 0.7--$3.5\kev$ energy band) are 
considerably lower. As we have discussed in Section 3, using the 
$L_X$--$\sigma$ correlation from Edge et al. suggests that clusters of the 
above X-ray luminosity should have velocity dispersions in the range 
400--$700\kms$ (cf., Figure~4), with an average value of $612\kms$.
Thus straightforward comparison with the present-day clusters shows that 
the average dispersion of the CEMM clusters is comparable with the value 
predicted on the basis of their space-density but higher than the value 
expected on the basis of X-ray luminosity. 

Such a direct comparison takes no account of the evolution of cluster 
properties expected between $z=0.41$ and the present-day. Only if the 
cosmological density parameter, $\Omega_0$, is extremely small can we hope 
to neglect the development of clusters between these epochs. To illustrate 
the expected evolution, we 
consider a family of models in which the universe has critical density (ie., 
$\Omega_0=1$) and the spectrum of density fluctuations can be adequately
described by a power-law on the scales relevant to galaxy clusters. 
For currently popular dark matter models, the power-law
index $n$ lies in the range $-1$ to $-1.5$; we consider $n=-1$ and $n=-2$ as
extremes. Under these conditions,
scaling relations (dependent on $n$ and an additional parameter
that describes the balance between heating and cooling of the
intra-cluster medium) can be used to predict the statistical properties of 
the cluster population at $z=0.4$ from the observed properties of
nearby clusters (eg., Kaiser, 1986, 1991, Evrard \& Henry, 1991 [EH], Bower, 
1997 [B97]).                                

\begintable{6}
{


  \def\ergs{\,\hbox{erg}\,\hbox{s}^{-1}}
  \def\kev{\,\hbox{keV}}

      \newdimen\digitwidth
      \setbox0=\hbox{\rm0}
      \digitwidth=\wd0
      \catcode`~=\active
      \def~{\kern\digitwidth}

	\def\-{\hbox{\ ---\ }}
	\def\:{\rlap{:}}
	\def\d{\rlap{$^\dagger$}}

	\def\kms{\,\hbox{km}\,\hbox{s}^{-1}}

      \def\mkrule{\noalign{\vskip 5pt}\noalign{\hrule}\noalign{\vskip 10pt}}

      \def\note#1{\item{$^{#1}$}}

\baselineskip=10pt
\tabskip 1em plus 2em minus 1em 

\noindent{\bf Table 6.} Predicted Velocity Dipersions

$$\vbox{
\halign{#\hfil&\quad#\hfil&\quad#\hfil&\quad#\hfil\cr
Evolutionary& \omit Spectral\hfil&  	\omit Entropy\hfil&  
						\omit\hfil$\sigma_v\,^a$\hfil\cr
Model&	      \omit Index ($n$)\hfil&	\omit Parameter ($\epsilon$)\hfil&
				\omit\hfil($\hbox{km}\,\hbox{s}^{-1}$)\hfil\cr
\mkrule
No Evolution&	\-&	\-&		612\cr
	\noalign{\vskip 10pt}
Self-Similar&	$-1$&	($-3$)&		515\cr
	&	$-2$&	($-5$)&		426\cr
	\noalign{\vskip 10pt}
Entropy Model&	$-1$&	~0&		579\cr
	&	$-1$&	+2&		625\cr
	&	$-2$&	~0&		517\cr
	&	$-2$&	+2&		559\cr
\mkrule
}
}$$

\leftline{Notes:}
\note{a} Mean velocity dispersion predicted for a cluster at redshift 0.41 with
	$L_X = 0.4 \times 10^{44} \ergs$ in the 0.7--$3.5 \kev$ energy band.
 
 }
\endtable

Table~6 compares the velocity dispersion of the composite dataset with the 
$L_X$--$\sigma$ correlation predicted in a number of evolutionary
models. The evolution of the correlation is calculated using the
appropriate scaling, and is then used to convert the average CEMM cluster X-ray 
luminosity into a predicted velocity 
dispersion.  In the first model, we consider the velocity dispersion 
expected if both the cluster virial temperature and X-ray luminosity 
evolve self-similarly (ie., the characteristic densities of  
the dark matter and the intra-cluster medium [ICM] both remain linked to the
background cosmological density). The 
predicted $L_X$--$\sigma$ correlation falls well below the composite 
CEMM data-point: for clusters of X-ray luminosity as low as is observed for 
the CEMM systems, the velocity dispersion is predicted to be of order 
$500\kms$. Altering the spectral index of fluctuations speeds up the
rate of evolution but fails to improve the comparison
since the reduction of the model clusters' X-ray luminosities is 
achieved only at the expense of also lowering the clusters' velocity dispersions. 

A greater range of evolutionary scenarios can be generated if we drop the 
assumption that the ICM follows the same evolution
as the dissipationless dark matter component.  A promising 
approach is to assume the intra-cluster gas has constant central entropy 
(ie., there is no net heating or cooling of the
ICM as the cluster evolves). Since the cluster's X-ray luminosity is very
sensitive to the central gas density, such a model naturally accounts
for the low X-ray luminosities of the clusters. The $L_X$--$\sigma$ 
correlation, however, remains close to its present-day
position. Thus, while this model achieves an improvement
over the self-similar models discussed above, the discrepancy between
the low X-ray luminosities and the high velocity dispersions of the
CEMM clusters remains a puzzle. In order to resolve this discrepancy,
it is necessary to increase the central entropy of the gas in the 
distant clusters relative to their present-day counterparts. This can be
described through an extension of EH's model that is described
in B97: the balance between heating and cooling of the ICM 
(and thus the evolution of its entropy) is described with
an additional parameter, $\epsilon$. The case $\epsilon=0$ 
corresponds to EH's original model, while for $\epsilon < 0$ shock
heating dominates the evolution of the ICM entropy, and for $\epsilon > 0$ 
the evolution is dominated by cooling. Closest agreement between 
the model and the observed properties of the CEMM clusters is obtained
by making $\epsilon$ positive. However, the cooling rate cannot be
made arbitrarily large: the dominant cooling mechanism is
X-ray radiation --- one of the quantities that we have measured both in the
local universe and at high redshift. B97 shows that a conservative upper 
limit is $\epsilon < 2$. When this constraint is included,
even this model fails to reproduce the positions of the CEMM clusters in 
the $(L_X,\sigma)$ plane, since the evolution of X-ray luminosity and 
velocity dispersion remain strongly coupled.


\subsection{A Non-Virial Interpretation of the CEMM Cluster Velocity Dispersions}

The difficulty in providing a theoretical explanation of the position of 
the single component fit in the $L_X$--$\sigma$ plane ensures that we  
consider alternative interpretations of the high overall dispersion. 
An artificially high dispersion might arise from (i) a number of distinct, similar
sized gravitational units superposed along the line of sight (the dispersion
within the individual systems makes them inseparable
in redshift space), or (ii) an infalling galaxy population that is bound to the
cluster but is not yet virialised (the dispersion of this component is inflated
by both the predominance of radial orbits and the physical separation).  

In order to address this possibility, we have investigated 
the robustness of our calculated dispersions to the exclusion of extreme 
members, and searched for evidence of a non-Gaussian velocity distribution.
When the clusters taken individually, there is no statistical evidence of 
substructure. There is, however, a visual impression of a core--halo
structure in many of the clusters (the notable exception being
F1637.23TL). Since this is very unlikely to occur by the overlap
of physically distinct systems, the infall model is to be preferred. 

By stacking all the clusters together, it is possible to quantify
the structure of the average velocity distribution.
The fit to this composite dataset is indeed improved if a low dispersion
centralised component ($\sim 200\kms$) is added to the high dispersion
($\sim 1000\kms$) component. The central component contains a
minority ($\sim 30\%$) of the cluster galaxies, however, suggesting that
it is better to interpret this as evidence for a leptokurtic (strongly peaked)
velocity distribution, rather than attaching definite physical
significance to the two components. Physically, such a line of sight 
velocity distribution indicates that a large fraction of the galaxies
are moving on radial orbits (eg., Merritt, 1987, Merritt \& Tremblay,
1994), a situation that is characteristic of a cluster that has not yet
reached virial equilibrium (ie., in which the galaxies' bulk motion
towards the cluster centre dominates over their random motion, Lynden-Bell
1967, Padmanabhan 1993).

One possibility to explain the anomalous dispersion of the CEMM clusters
is therefore that these clusters are 
seen at a period of rapid growth. This picture is appealing since evolution in
the dynamical state of clusters might be linked to changes in their galaxy
populations (Kauffmann, 1995): specifically, the increased fraction of blue galaxies
(eg., Butcher \& Oemler, 1978) and the increased proportion of galaxies showing
spectral evidence of recent or on-going intense bursts of star formation
(eg., Dressler \& Gunn, 1983, Couch \& Sharples, 1987, Barger et al., 1996). 
However, another possibility is
that the clusters are embedded in large-scale filaments that are oriented
along the line of sight. Such systems may have been preferentially selected
by the catalogue's optical selection criterion. This effect could boost
the apparent infall rate in these systems, making them poorly representative
of the distant cluster population as a whole.

In the latter case, we would expect that low redshift optically selected 
clusters would occupy a similar position in the $L_X$--$\sigma$ diagram.
To investigate this, we first used clusters from the Zabludoff et al., 1990,
survey for which X-ray data was available from the David et al. (1994)
compilation. These data points are shown as open triangles in Figure~4.
The two lowest X-ray luminosity clusters 
($L_X < 0.5\times10^44 \ergs$), indeed lie above our fiducial correlation.
Similar results are found for a larger sample of clusters taken from the 
Edinburgh/Durham Cluster Catalogue (EDCC, Collins et al., 1995,
Nichol et al., in prep.). It is not clear whether these data-sets imply a 
flattening of the $L_X$--$\sigma$ correlation at low luminosities or 
whether averaging over sufficient clusters would produce a mean
point lying on the fiducial correlation. Nevertheless, there is clearly
evidence for a population of low X-ray luminosity, high velocity 
dispersion systems at low redshift. These are the nearby counterparts
of the CEMM clusters.  The existence of such a population 
in optically selected cluster samples suggests that the selection
process cannot be separated from the properties of the resulting cluster
population. A connection between the anomalous velocity dispersions and
the Butcher-Oemler effect cannot yet be ruled out, however, since we know little 
about the stellar populations of the galaxies in the nearby systems.

\section{Conclusions}

In this paper, we set out to explore the dynamics of a representative 
sample of galaxy clusters taken from the Couch et al.\ (1991, CEMM)
catalogue. The catalogue identifies clusters as enhancements in the
densities of faint objects on deep 4-m optical plates, with initial
spectroscopic work being undertaken by CEMM to establishing
cluster redshifts on the basis of between 2 and 3 concordant 
galaxies. Following up this catalogue at X-ray wavelengths,
we showed (Bower et al., 1994, Paper~I) that the 
X-ray luminosities of these clusters were well below those of 
present-day clusters of comparable space density. This is an important
result as it reinforces the decline in the amplitude of the X-ray 
luminosity function of galaxy clusters that is seen in X-ray selected 
samples (Henry et al., 1992, Castander et al., 1995). 

Here, we have undertaken further spectroscopic work in order
to increase the number of galaxy redshifts that are available in 
seven of these clusters. This programme
is designed to investigate the dynamics of the clusters and to compare
the clusters' velocity dispersions with their X-ray luminosities.

Firstly, however, the study answers a possible criticism of Paper~I by 
establishing the extent to which the optically selected
catalogue is contaminated by {\it projected\/}
galaxy over-densities that result from the superposition of several
poor clusters or groups along the line of sight. Such projected associations
would not be expected to be X-ray luminous. Typically, we 
have increased the number of known cluster members to more than 10. In all
but one case, the extended redshift survey has confirmed the physical
association of the galaxies, a well-defined peak in the redshift distribution
being clearly visible. 

The main focus of this paper is the velocity distribution of the galaxies in 
these clusters. This is characterised by the dispersion about the cluster
mean. If we make the assumption that these systems are relaxed, 
it is valid to use the spread of the cluster members' redshifts in order
to estimate the depth of the clusters' gravitational potential. 
Comparison of this with the clusters X-ray luminosities allows us to 
investigate the efficiency of the thermal plasma emission
in a way that is independent of the cluster's space-density. This 
approach thus provides an important comparison with the results of Paper~I. 

We have shown that the clusters' velocity dispersion is high: comparable with
that expected on the basis of the clusters' space-density but much higher
than would be expect on the basis of the clusters'
X-ray luminosity. This conclusion applies whether we consider
the clusters individually, or as a composite system.

Relative to the clusters' X-ray luminosities, these dispersions appear to
demonstrate that considerable evolution has taken place in the 
X-ray properties of clusters since $z=0.4$. However, the strength of 
the evolution does not match any simple theoretical expectation.
The rate of evolution in the X-ray luminosity
-- virial temperature correlation that is implied cannot be reproduced
even within an extreme model of cluster evolution dominated by radiative cooling.  
Thus although our results seem initially to be in qualitative agreement with 
observations of a decline in the X-ray luminosities of galaxy clusters,
the strength of the effect measured seems unreasonably large. 

The lack of a simple explanation for the position of the clusters in the 
$L_X$--$\sigma$ plane forces us to question whether the velocity dispersions
we derive can be indicative of the clusters' masses. Closer examination
of the velocity histograms of several of the clusters gives an impression
that the distribution is excessively peaked compared to a Gaussian, 
suggesting a core/halo structure. Although this 
impression is not statistically significant in the individual clusters,
it is confirmed at a high level of significance in the analysis of the 
stacked cluster dataset. The composite cluster is well described by
a core component of dispersion $\sim 200\kms$,
and a halo of dispersion $\sim 1000\kms$. However, the halo component 
accounts for the larger part of the galaxy population, the core
accounting for less than 30\% of the total. This suggests that the separate components
probably do not have a distinct physical significance; more likely they 
simply reflect the peakedness of the velocity distribution.  

Our data therefore suggest that the galaxy population of these clusters is 
dominated by an infalling galaxy halo. We have put forward two possible explanations, 
one centered on the evolution of clusters, the other resulting from a 
selection bias. The first interpretation is that these clusters that 
have not yet formed --- or at least are caught at a time of rapid
mass growth.  The second possibility appears more relevant, however:
potentially the clusters selected by CEMM are embedded in large scale 
structure `filaments' that are being viewed `end-on'. Rather than being 
a consequence of cluster evolution, this could plausibly result from the 
clusters' optical selection criterion. This interpretation is given support 
by the existence of similar low X-ray luminosity - high velocity dispersion 
systems in optically selected surveys at low redshift.
Such an effect might explain the presence of an excess blue galaxy 
population in some distant clusters, although we would expect the 
additional galaxies to be drawn from the general field galaxy population
rather than being peculiar to the cluster environment.
What remains uncertain, however, is whether an evolutionary connection
is required: do our results imply an increase in the population of
anomalous high dispersion clusters with increasing redshift? Unfortunately,
our catalogue is too small and may not adequately sample the full
distant cluster population. In order to investigate evolution in optically 
selected cluster samples, larger cluster catalogues --- with matched 
selection criteria --- are needed at both high and low redshift.

\section*{Acknowledgments}

We are happy to thank ESO for the provision of the telescope time
that made this project possible, and to acknowledge the Starlink project
for the provision of computing support. We thank Sharon Lippey for her
assistance with the redshift measurements as part of her undergraduate work
at the UNSW, and the referee, Bob Nichol, for his helpful comments and
suggestions.

\section*{Finding Charts}

{\it A compressed tar file containing the full set of finding charts can be
obtained from {\sl http://star-www.dur.ac.uk/~rgb/}}.

\smallskip

\noindent
{\bf Plate 1(a--g).} Finding charts of the CEMM cluster fields studied in this
	project: (a)~F1557.19TC, (b)~F1652.20CR, (c)~F1637.23TL, 
	(d)~J2175.15TR, (e)~J2175.23C,\break (f)~F1835.22CR, (g)~F1835.2CL. 
	Objects are labeled either
	by their original CEMM designation, or by the slit number with prefix A
	or B to distinguish the first and second masks. All the images
	are reproduced to the same angular scale, with the larger images
	giving a $5'\times5'$ area. For $H_0=50\kms\Mpc^{-1}$ and $q_0=0.5$,
	this corresponds to a $2\times2\Mpc^2$ region at $z=0.41$.
	Our spectroscopy concentrated
	on objects close to the cluster center, however: almost all cluster
	members come from within the central $0.5 \Mpc$.
	With the exception of Plate~1f (F1835.22CR), the field is oriented
	with North up and East to the left; Plate~1f is rotated by $45^\circ$
	so that the top right-hand corner points North. 

\bye